\def\etal{{\it et al.}}
\def\eg{{\it e.g.}}
\def\ie{{\it i.e.}}

\def\calo{{\cal O}}
\def\calm{{\cal M}}

\def\theories{{\cal Q}}

\def\redef{{\cal G}}
\def\fields{{\cal F}}

\def\crit{{\cal C}}
\def\gmn{g_{\mu\nu}}
\documentclass[cup7b]{cupbook}
\usepackage{amsmath}
\usepackage{graphicx}
\begin{document}

\centerline{\bf \huge Asymptotic Safety}
\medskip
\centerline{\bf Roberto Percacci}
\smallskip
\centerline{SISSA, via Beirut 4, 34014, Trieste, Italy}
\centerline{INFN, Sezione di Trieste}

{\large 
\vskip 4cm
{\bf Abstract.} Asymptotic safety is a set of conditions, 
based on the existence of a nontrivial fixed point for the renormalization group flow, 
which would make a quantum field theory consistent up to arbitrarily high energies.
After introducing the basic ideas of this approach, I review the present evidence 
in favor of an asymptotically safe quantum field theory of gravity.
\vskip 1cm
\centerline{To appear in `` Approaches to Quantum Gravity:}
\centerline{Towards a 
New Understanding of Space, Time and Matter''} 
\centerline{ed. D. Oriti, Cambridge University Press.}
}
\vfil
\break

\author[R. Percacci]{R. PERCACCI\\SISSA, Trieste}
\chapter{Asymptotic Safety} 
\begin{abstract}

Asymptotic safety is a set of conditions, 
based on the existence of a nontrivial fixed point for the renormalization group flow, 
which would make a quantum field theory consistent to arbitrarily high energies.
After introducing the basic ideas of this approach, I review the present evidence 
in favor of an asymptotically safe quantum field theory of gravity.
\end{abstract}

\section{Introduction}
The problems of perturbative Quantum Field Theory (QFT) in 
relation to the UV behaviour of gravity
have led to widespread pessimism about the possibility of constructing
a fundamental QFT of gravity.
Instead, we have become accustomed to thinking of General Relativity
(GR) as an effective field theory, which only gives an accurate 
description of gravitational physics at low energies.
The formalism of effective field theories provides a coherent framework 
in which quantum calculations can be performed even if the theory is not renormalizable.
For example, quantum corrections to the gravitational potential
have been discussed by several authors; see 
Bjerrum-Bohr \etal\ (2003) and references therein.
This continuum QFT description is widely expected to break down at very short
distances and to be replaced by something dramatically different
beyond the Planck scale.
There is however no proof that continuum QFT will fail, and the current situation
may just be the result of the lack of suitable technical tools.
Weinberg (1979) described a generalized, nonperturbative notion
of renormalizability called ``asymptotic safety'' and suggested that
GR may satisfy this condition, making it a consistent QFT at all energies.
The essential ingredient of this approach is the existence of a Fixed Point (FP)
in the Renormalization Group (RG) flow of gravitational couplings.
Several calculations were performed using the $\epsilon$--expansion around $d=2$ 
dimensions, supporting the view that gravity is asymptotically safe
(Gastmans \etal\ (1978), Christensen \& Duff (1978), Kawai \& Ninomiya (1990)).
However, the continuation to four dimensions ($\epsilon\to 2$) 
was questionable and this line of research slowed down for some time.
It was revived by Reuter (1998) who calculated the gravitational
beta functions directly in $d=4$ dimensions, using a truncation of an
Exact Renormalization Group Equation (ERGE).
Matter couplings were considered by Dou \& Percacci (1998);
then Souma (1999) found that these beta functions admit a non--Gau\ss ian FP.
Further work by Lauscher \& Reuter (2002a,b), Percacci (2006), Codello \& Percacci (2006)
strongly supports the view that this FP is not a mere artifact of the approximations made.
An extensive review of this subject can be found in Niedermaier \& Reuter (2006).

In section 1.2 I introduce the general idea of asymptotic safety;
the reader is referred to Weinberg (1979) for a more detailed discussion.
In section 1.3 I describe some peculiarities of the gravitational RG,
which derive from the dual character of the metric as a dynamical field
and as definition of lengths.
Recent evidence for a FP, coming mainly from the ERGE, is reviewed in section 1.4.
Some relations to other approaches to quantum gravity 
are briefly mentioned in section 1.5.

\medskip
\section{The general notion of asymptotic safety}
\smallskip
The techniques of effective QFT have been recognized as being
of great generality and are now quite pervasive in particle physics.
An effective field theory is described by an effective action  $\Gamma_k$
which can be thought of as the result of having integrated out all fluctuations 
of the fields with momenta larger than $k$.
We need not specify here the physical meaning of $k$:
for each application of the theory one will have to identify the
physically relevant variable acting as $k$
(in particle physics it is usually some external momentum).
One convenient definition of $\Gamma_k$ that we shall use here is as follows.
We start from a (``bare'') action $S[\phi_A]$ 
for multiplets of quantum fields $\phi_A$, 
describing physics at an energy scale $k_0$.
We add to it a term $\Delta S_k[\phi_A]$, quadratic in the $\phi_A$, 
which in Fourier space has the form:
$\Delta S_k[\phi]=\int d^d q \phi_A R_k^{AB}(q^2)\phi_B$.
The kernel $R_k^{AB}(q^2)$, henceforth called the cutoff function,
is chosen in such a way that the propagation of
field modes $\phi_A(q)$ with momenta $q<k$ is suppressed, while
field modes with momenta $k<q<k_0$ are unaffected.
We formally define a $k$--dependent generating functional of connected Green functions 
$$
W_k[J^A]=-\log \int (d\phi_A) \exp\left(-S[\phi_A]-\Delta S_k[\phi_A]-\int J^A\phi_A\right)
\eqno(1.2.1)
$$
and a modified $k$--dependent Legendre transform
$$
\Gamma_k[\phi_A]=W_k[J^A]-\int J^A\phi_A -\Delta S_k[\phi_A]\ ,\eqno(1.2.2)
$$
where $\Delta S_k$ has been subtracted.
The ``classical fields'' $\frac{\delta W_k}{\delta J^A}$
are denoted again $\phi_A$ for notational simplicity.
This functional interpolates continuously between $S$, for $k=k_0$, 
and the usual effective action $\Gamma[\phi_A]$,
the generating functional of one--particle irreducible Green functions, for $k=0$.
It is similar in spirit, but distinct from, the Wilsonian effective action.
In the following we will always use this definition of $\Gamma_k$,
but much of what will be said should be true also with other definitions.

In the case of gauge theories there are complications due to the fact that
the cutoff interferes with gauge invariance. 
One can use a background gauge condition, which circumvents these
problems by defining a functional of two fields, the background field and the classical field;
the effective action $\Gamma_k$ is then obtained by identifying these fields.
See Pawlowski (2005), or Reuter (1998) for the case of gravity.

The effective action $\Gamma_k[\phi_A]$, used at tree level, 
gives an accurate description of processes occurring at momentum scales of order $k$.
In general it will have the form
$\Gamma_k(\phi_A,g_i)=\sum_i g_i(k) {\cal O}_i(\phi_A)$,
where $g_i$ are runnning coupling constants and 
${\cal O}_i$ are all possible operators constructed with the fields
$\phi_A$ and their derivatives, which are compatible 
with the symmetries of the theory.
It can be thought of as a functional on $\fields\times\theories\times R^+$,
where $\fields$ is the configuration space of the fields, $\theories$
is an infinite dimensional manifold parametrized by the coupling constants,
and $R^+$ is the space parametrized by $k$.
The dependence of $\Gamma_k$ on $k$ is given by
$\partial_t\Gamma_k(\phi_A,g_i)=\sum_i \beta_i(k) {\cal O}_i(\phi_A)$
where $t=\log(k/k_0)$ and
$\beta_i(g_j,k)=\partial_t g_i$ are the beta functions.

Dimensional analysis implies the scaling property
$$
\Gamma_k(\phi_A,g_i)=\Gamma_{bk}(b^{d_A} \phi_A ,b^{d_i}  g_i )\ ,\eqno(1.2.3)
$$
where $d_A$ is the canonical dimension of $\phi_A$,
$d_i$ is the canonical dimension of $g_i$, and $b\in R^+$
is a positive real scaling parameter
\footnote{We assume that the coordinates are dimensionless,
as is natural in curved space,
resulting in unconventional canonical dimensions.
The metric is an area.}.
One can rewrite the theory in terms of dimensionless fields 
$\tilde\phi_A=\phi_A k^{-d_A}$ and dimensionless couplings $\tilde g_i=g_i k^{-d_i}$.
A transformation (1.2.3) with parameter $b=k^{-1}$ can be used to
define a functional $\tilde\Gamma$ on $(\fields\times\theories\times R^+)/R^+$:
$$
\tilde\Gamma(\tilde\phi_A,\tilde g_i):=
\Gamma_{1}(\tilde\phi_A ,\tilde g_i )=
\Gamma_k(\phi_A,g_i)\ .
\eqno(1.2.4)
$$
Similarly, $\beta_i(g_j,k)=k^{d_i}a_i(\tilde g_j)$
where $a_i(\tilde g_j)=\beta_i(\tilde g_j,1)$.
There follows that the beta functions of the dimensionless couplings,
$$
\tilde\beta_i(\tilde g_j)\equiv\partial_t \tilde g_i=a_i(\tilde g_j)-d_i\tilde g_i
\eqno(1.2.5)
$$
depend on $k$ only implicitly via the $\tilde g_j(t)$.

The effective actions $\Gamma_k$ and $\Gamma_{k-\delta k}$
differ essentially by a functional integral over field modes 
with momenta between $k$ and $k-\delta k$.
Such integration does not lead to divergences, so the beta functions 
are automatically finite.
Once calculated at a certain scale $k$, they are automatically determined 
at any other scale by dimensional analysis.
Thus, the scale $k_0$ and the ``bare'' action $S$ act just as initial conditions:
when the beta functions are known,
one can start from an arbitrary initial point on $\theories$ 
and follow the RG trajectory in either direction.
The effective action $\Gamma_k$ at any scale $k$ can be obtained integrating the flow. 
In particular, the UV behaviour can be studied by taking the limit $k\to\infty$.

It often happens that the flow cannot be integrated beyond
a certain limiting scale $\Lambda$, defining the point 
at which some ``new physics'' has to make its appearance. 
In this case the theory only holds
for $k<\Lambda$ and is called an ``effective'' or ``cutoff'' QFT.
It may happen, however, that the limit $t\to\infty$
can be taken; we then have a self-consistent description of a certain set
of physical phenomena which is valid for arbitrarily high energy scales
and does not need to refer to anything else outside it.
In this case the theory is said to be ``fundamental''.

The couplings appearing in the effective action can be related
to physically measurable quantities such as cross
sections and decay rates. Dimensional analysis implies that aside
from an overall power of $k$, such quantities only depend on
dimensionless kinematical variables $X$, like scattering angles and ratios
of energies, and on the dimensionless couplings $\tilde g_i$ 
(recall that usually $k$ is identified with one of the momentum variables).
For example, a cross section can be expressed as 
$\sigma=k^{-2}\tilde\sigma(X,\tilde g_i)$.
If some of the couplings $\tilde g_i$ go to infinity 
when $t\to\infty$, also the function $\tilde\sigma$ can be expected to diverge.
A sufficient condition to avoid this problem is to assume that
in the limit $t\to\infty$ the RG trajectory tends to
a FP of the RG, \ie\ a point $\tilde g_*$ 
where $\tilde\beta_i(\tilde g_{*})=0$ for all $i$.
The existence of such a FP is the first requirement for asymptotic safety.
Before discussing the second requirement, we have to understand
that one needs to impose this condition only on a subset of all couplings.

The fields $\phi_A$ are integration variables, and a redefinition 
of the fields does not change the physical content of the theory.
This can be seen as invariance under a group $\redef$ of coordinate
transformations in $\fields$.
There is a similar arbitrariness in the choice of coordinates
on $\theories$, due to the freedom of redefining the couplings $g_i$.
Since, for given $k$, $\Gamma_k$ is assumed to be the ``most general''
functional on $\fields\times\theories$ (in some proper sense),
given a field redefinition $\phi'=\phi'(\phi)$
one can find new couplings $g'_i$ such that
$$
\Gamma_k(\phi'_B(\phi_A),g_i)=\Gamma_k(\phi_A,g'_i)\ .\eqno(1.2.6)
$$
At least locally, this defines an action of $\redef$ on $\theories$.
We are then free to choose a coordinate system
which is adapted to these trasformations, in the sense that
a subset $\{g_{\hat \imath}\}$ of couplings transform nontrivially 
and can be used as coordinates in the orbits of $\redef$,
while a subset $\{g_{\bar \imath}\}$ are invariant under the action
of $\redef$ and define coordinates on $\theories/\redef$.
The couplings $g_{\hat \imath}$ are called redundant or inessential,
while the couplings $g_{\bar \imath}$ are called essential.
In an adapted parametrization there exists, at least locally, 
a field redefinition $\bar\phi(\phi)$ such that using (1.2.6)
the couplings $g_{\hat \imath}$ can be given fixed values $(g_{\hat \imath})_0$.
We can then define a new action $\bar\Gamma$ depending only on the
essential couplings:
$$
\bar\Gamma_k(\bar\phi_A,g_{\bar \imath}):=
\Gamma_k(\bar\phi_A,g_{\bar \imath},(g_{\hat \imath})_0)
=\Gamma_k(\phi_A;g_{\bar \imath},g_{\hat \imath})\ .\eqno(1.2.7)
$$
Similarly, the values of the redundant couplings can be fixed also
in the expressions for measurable quantities, 
so there is no need to constrain their RG flow in any way:
they are not required to flow towards a FP.

For example, the action of a scalar field theory in a background $\gmn$,
$$
\Gamma_k(\phi,g_{\mu\nu};Z_\phi,\lambda_{2i})=\int d^4x\sqrt{g}\left[
{\frac{Z_\phi}{2}}g^{\mu\nu}\partial_\mu\phi\partial_\nu\phi
+\lambda_2\phi^2+\lambda_4\phi^4+\ldots
\right]\ \eqno(1.2.8)
$$
has the scaling invariance
$$
\Gamma_k(c\phi,g_{\mu\nu};c^{-2}Z_\phi,c^{-2i}\lambda_{2i})=
\Gamma_k(\phi,g_{\mu\nu};Z_\phi,\lambda_{2i})\ ,\eqno(1.2.9)
$$
which is a special case of (1.2.6).
There exists an adapted coordinate system where $Z$ is inessential
and $\bar\lambda_{2i}=\lambda_{2i}Z_\phi^{-i}$
are the essential coordinates.
A transformation with $c=\sqrt{Z_\phi}$ then leads to $Z_\phi=1$,
leaving the essential couplings unaffected.

A comparison of (1.2.4) and (1.2.7) shows that $k$ behaves like 
a redundant coupling.
In ordinary QFT's, it is generally the case that for each multiplet 
of fields $\phi_A$ there is a scaling invariance like (1.2.9)
commuting with (1.2.3). One can use these invariances to eliminate
simultaneously $k$ and one other redundant coupling per field multiplet; 
the conventional choice is to eliminate the wave function 
renormalization $Z_A$.
No conditions have to be imposed on the RG flow of the $Z_A$'s, 
and the anomalous dimensions $\eta_A=\partial_t\log Z_A$,
at a FP, can be determined by a calculation.
More generally, (1.2.3) and (1.2.6) can be used 
to eliminate simultaneously the dependence of $\Gamma_k$ on $k$ and
on the inessential couplings, and to define an effective action
$\tilde\Gamma(\tilde\phi_A,\tilde g_{\bar \imath})$, depending only on the
dimensionless essential couplings 
$\tilde g_{\bar \imath}=g_{\bar \imath}k^{-d_{\bar\imath}}$.
It is only on these couplings that one has to impose the FP
condition $\partial_t\tilde g_{\bar \imath}=0$.

We can now state the second requirement for asymptotic safety.
Denote $\tilde\theories=(\theories\times R^+)/(\redef\times R^+)$
the space parametrized by the dimensionless essential couplings 
$\tilde g_{\bar \imath}$.
The set $\crit$ of all points in $\tilde\theories$ that flow towards the 
FP in the UV limit is called the UV critical surface.
If one chooses an initial point lying on $\crit$, the whole trajectory
will remain on $\crit$ and will ultimately flow towards the FP
in the UV limit.
Points that lie outside $\crit$ will generally flow towards infinity
(or other FP's).
Thus, demanding that the theory lies on the UV critical surface
ensures that it has a sensible UV limit.
It also has the effect of reducing the arbitrariness in the choice of the coupling constants.
In particular, if the UV critical surface is finite dimensional,
the arbitariness is reduced to a finite number of parameters,
which can be determined by a finite number of experiments.
Thus, a theory with a FP and a finite dimensional UV 
critical surface has a controllable UV behaviour, and is predictive.
Such a theory is called ``asymptotically safe''.

A perturbatively renormalizable, asymptotically free field theory
such as QCD is a special case of an asymptotically safe theory. 
In this case the FP is the Gau\ss ian FP, where
all couplings vanish, and the critical surface is spanned,
near the FP, by the couplings that are renormalizable in the perturbative sense
(those with dimension $d_{\bar\imath}\geq 0$).

The requirement of renormalizability played an important role
in the construction of the Standard Model (SM) of particle physics.
Given that the SM is not a complete theory, and that some of its
couplings are not asymptotically free, nowadays it is
regarded an effective QFT, whose nonrenormalizable couplings
are suppressed by some power of momentum over cutoff.
On the other hand, any theory that includes both the SM and gravity 
should better be a fundamental theory.
For such a theory, the requirement of asymptotic safety will have the same 
significance that renormalizability originally had for the SM.

\medskip
\section{The case of gravity}
\smallskip
We shall use a derivative expansion of $\Gamma_k$:
$$
\Gamma_k(g_{\mu\nu};g^{(n)}_i)=
\sum_{n=0}^\infty \sum_i g^{(n)}_i(k) {\cal O}^{(n)}_i(g_{\mu\nu})\ ,\eqno(1.3.1)
$$
where $\calo^{(n)}_i=\int d^dx\,\sqrt{g}\calm^{(n)}_i$ and $\calm^{(n)}_i$
are polynomials in the curvature tensor and its derivatives
containing $2n$ derivatives of the metric; $i$ is an index that
labels different operators with the same number of derivatives.
The dimension of $g^{(n)}_i$ is $d_n=d-2n$.
The first two polynomials are just $\calm^{(0)}=1$, $\calm^{(1)}=R$.
The corresponding couplings are
$g^{(1)}=-Z_g=-\frac{1}{16\pi G}$, 
$g^{(0)}=2 Z_g\Lambda$, $\Lambda$ being the cosmological constant.
Newton's constant $G$ appears in $Z_g$, which in linearized Einstein theory
is the wave function renormalization of the graviton.
Neglecting total derivatives, one can choose as terms with
four derivatives of the metric 
$\calm^{(2)}_1=C^2$ (the square of the Weyl tensor) and
$\calm^{(2)}_2=R^2$.
We also note that the coupling constants of higher derivative gravity
are not the coefficients $g^{(2)}_i$ but rather their inverses $2\lambda=(g^{(2)}_1)^{-1}$ and 
$\xi=(g^{(2)}_2)^{-1}$.
Thus,
$$
\Gamma_k^{(n\leq 2)}=\int d^dx\,\sqrt{g}\left[
2 Z_g\Lambda-Z_gR+\frac{1}{2\lambda}C^2+
\frac{1}{\xi}R^2\right]\ .\eqno(1.3.2)
$$
As in any other QFT, $Z_g$ can be eliminated from the
action by a rescaling of the field.
Under constant rescalings of $g_{\mu\nu}$, in $d$ dimensions,
$$
\Gamma_k(g_{\mu\nu};g^{(n)}_i)=
\Gamma_{bk}(b^{-2}g_{\mu\nu};b^{d-2n}g^{(n)}_i)\ .\eqno(1.3.3)
$$
This relation is the analog of (1.2.9) for the metric,
but also coincides with (1.2.3), the invariance at the basis of dimensional analysis;
fixing it amounts to a choice of unit of mass.
This is where gravity differs from any other field theory
(Percacci \& Perini (2004), Percacci (2007)).
In usual QFT's such as (1.2.8), one can exploit the two invariances (1.2.3) and (1.2.9) 
to eliminate simultaneously $k$ and $Z$ from the action.
In the case of pure gravity there is only one such invariance and
one has to make a choice.

If we choose $k$ as unit of mass, we can define the effective action,
$$
\tilde\Gamma(\tilde g_{\mu\nu};\tilde Z_g,\tilde\Lambda,\ldots)=
\Gamma_1(\tilde g_{\mu\nu};\tilde Z_g,\tilde\Lambda,\ldots)=
\Gamma_k(\gmn;Z_g,\Lambda,\ldots)\ ,\eqno(1.3.4)
$$
where $\tilde g_{\mu\nu}=k^{2}g_{\mu\nu}$,
$\tilde Z_g=\frac{Z_g}{k^2}=\frac{1}{16\pi \tilde G}$,
$\tilde\Lambda=\frac{\Lambda}{k^2}$, etc..
There is then no freedom left to eliminate $Z_g$.
Physically measurable quantities will depend explicitly on $\tilde Z_g$,
so by the arguments of section 1.2, we have to impose
that $\partial_t \tilde Z_g=0$, or equivalently $\partial_t \tilde G=0$, at a FP.

Alternatively, one can use (1.3.3) to set $Z_g=1$:
this amounts to working in Planck units.
Then we can define a new action
\footnote{Note that to completely eliminate $Z_g$ from the action
one has to scale the whole metric, and not just the fluctuation, as is customary in perturbation theory.}:
$$
\Gamma'_{k'}(g'_{\mu\nu};\Lambda',\ldots)=
\Gamma_{k'}(g'_{\mu\nu};\Lambda',1,\ldots)=
\Gamma_k(\gmn;\Lambda,Z_g,\ldots)\ ,
\eqno(1.3.5)
$$
where $g'_{\mu\nu}=16\pi Z_g\gmn$,
$\Lambda'=\frac{1}{16\pi Z_g}\Lambda$,
$k'=\sqrt{\frac{1}{16\pi Z_g}}k$ etc.
are the metric, cosmological constant and cutoff measured in Planck units.
In this case, the dependence on $G$ disappears; 
however, the beta functions and measurable quantities
will depend explicitly on $k'$.

In theories of gravity coupled to matter, the number of these scaling invariances
is equal to the number of field multiplets, so the situation is the same
as for pure gravity. (Without gravity, it is equal to the number of field
multiplets plus one, due to dimensional analysis.) 
The situation can be summarized by saying that when the metric
is dynamical, $k$ should be treated as one of the couplings, and that
there exist parametrizations where $k$ is redundant or $G$
is redundant, but not both.

Scale invariance is usually thought to imply that a theory contains
only dimensionless parameters, and the presence at a FP of nonvanishing 
dimensionful couplings may seem to be at odds with
the notion that the FP theory is scale--invariant.
This is the case if only the fields are scaled, and not the couplings.
In an asymptotically safe QFT, scale invariance is realized in another way:
all dimensionful couplings scale with $k$ as required by their canonical dimension.
In geometrical terms, the RG trajectories in $\theories$
lie asymptotically in an orbit of the transformations (1.2.3) and (1.2.6).
This also has another consequence.
At low momentum scales $p\ll \sqrt{Z_g}$ the couplings are not expected to run
and the terms in the action (1.3.2) with four derivatives are
suppressed relative to the term with two derivatives by a factor $p^2/Z_g$.
On the other hand in the FP regime, if we evaluate the couplings at $k=p$, 
the running of $Z_g$ exactly compensates the effect
of the derivatives: both terms are of order $p^4$. 
From this point of view, {\it a priori} all terms in (1.3.1) could be equally important. 

From the existence of a FP for Newton's constant there would immediately follow 
two striking consequences.
First, the cutoff measured in Planck units would be bounded.
This is because the cutoff in Planck units, $k'=k\sqrt{G}$, is equal to the square root of
Newton's constant in cutoff units, $\sqrt{\tilde G}$. 
Since we have argued that the latter must have a finite limit at a FP, 
then also the former must do so.
This seems to contradict the notion that the UV limit is defined by $k\to\infty$.
The point is that only statements about dimensionless quantities 
are physically meaningful, and the statement ``$k\to\infty$'' is 
meaningless until we specify the units. 
In a fundamental theory one cannot refer to any
external ``absolute'' quantity as a unit, and any internal quantity
which is chosen as a unit will be subject to the RG flow.
If we start from low energy ($k'\ll 1$) and we increase $k$, 
$k'$ will initially increase at the same rate, because in this regime
$\partial_t G\approx 0$; however, when $k'\approx 1$ we reach the FP regime 
where $G(k)\approx\tilde G_*/k^2$ and therefore $k'$ stops growing.

The second consequence concerns the graviton anomalous dimension, which in $d$
dimensions is $\eta_g=\partial_t \log Z_g=\partial_t \log\tilde Z_g+d-2$.
Since we have argued that $\partial_t\tilde Z_g=0$ at a gravitational FP,
if $\tilde Z_{g*}\not=0$ we must have $\eta_{g*}=d-2$.
The propagator of a field with anomalous dimension $\eta$ behaves like $p^{-2-\eta}$,
so one concludes that at a nontrivial gravitational FP the graviton propagator
behaves like $p^{-d}$ rather than $p^{-2}$, as would follow from a naive
classical interpretation of the Einstein-Hilbert action.
Similar behaviour is known also in other gauge theories away from
the critical dimension, see \eg\ Kazakov (2003).

\medskip
\section{The Gravitational Fixed Point}
\smallskip
I will now describe some of the evidence that has accumulated in favor of 
a nontrivial gravitational FP.
Early attempts were made in the context of the
$\epsilon$--expansion around two dimensions ($\epsilon=d-2$), which yields 
$$
\beta_{\tilde G}=\epsilon \tilde G-q \tilde G^2\ .
\eqno(1.4.1)
$$
Thus there is a UV--attractive FP at $\tilde G_*=\epsilon/q$.
The constant $q=\frac{38}{3}$ for pure gravity 
(Weinberg (1979), Kawai \& Ninomiya (1990),
see Aida \& Kitazawa (1997) for two--loop results).
Unfortunately, for a while it was not clear whether one could
trust the continuation of this result to four dimensions ($\epsilon=2$).

Most of the recent progress in this approach has come from the application 
to gravity of the ERGE.
It was shown by Wetterich (1993) that the effective action $\Gamma_k$ 
defined in (1.2.2) satisfies the equation
$$
\partial_t\Gamma_k=
\frac{1}{2}\mathrm{STr}\left(\frac{\delta^2\Gamma_k}{\delta\phi_A\delta\phi_B}+R_k^{AB}\right)^{-1}
\partial_tR_k^{BA}\ ,\eqno(1.4.2)
$$
where STr is a trace over momenta as well as over particle species
and any spacetime or internal indices,
including a sign -1 for fermionic fields and a factor 2 for complex fields.
In the case of gauge theories, the ghost fields have to be included among the $\phi_A$.

Comparing the r.h.s. of the ERGE with the $t$-derivative of (1.3.1)
one can extract the beta functions.
Note that in general the cutoff function $R_k$ may depend on the couplings and therefore
the term $\partial_t R_k$ in the r.h.s. of (1.4.2) contains the beta functions.
Thus, extracting the beta functions from the ERGE implies solving an equation
where the beta functions appear on both sides.
At one loop, the effective action $\Gamma_k$ is 
${\rm Tr}\log\frac{\delta^2 (S+\Delta S_k)}{\delta\phi\delta\phi}$;
it satisfies an equation which is formally identical to (1.4.2) except that in the r.h.s.
the running couplings $g_i(k)$ are replaced everywhere by the ``bare'' 
couplings $g_i(k_0)$, appearing in $S$.
We will call ``one--loop beta functions'' those extracted from the ERGE 
ignoring the derivatives of the couplings that may appear in the r.h.s. of (1.4.2).

It is usually impossible to get the beta functions for all couplings,
so a common procedure is to consider a truncation of the theory
where the effective action $\Gamma_k$ contains only a (finite or infinite) subset 
of all possible terms.
In these calculations there is no small parameter to tell us
what terms can be safely neglected, so the choice of truncation
has to be motivated by physical insight.
On the other hand, in this way one can obtain genuine nonperturbative information.
This and other similar ERGEs have been applied to a variety of problems.
One can reproduce the universal one loop beta functions of familiar theories,
and in more advanced approximations the results are quantitatively 
comparable to those obtainable by other methods.
See Bagnuls \& Bervilliers (2001), Berges \etal\ (2002), Pawlowski (2005)
for reviews.

The simplest way to arrive at a gravitational FP in four dimensions,
avoiding the technical complications of graviton propagators,
is through the contributions of matter loops
to the beta functions of the gravitational couplings.
Thus, consider gravity coupled to $n_S$ scalar fields,
$n_D$ Dirac fields, $n_M$ gauge (Maxwell) fields, all massless and minimally coupled.
A priori, nothing is assumed about the gravitational action.
For each type of field $\phi_A$ we choose the cutoff function in such a way that
$P_k(\Delta^{(A)})=\Delta^{(A)}+R_k(\Delta^{(A)})$, 
where $\Delta^{(S)}=-\nabla^2$ on scalars, $\Delta^{(D)}=-\nabla^2+\frac{R}{4}$  on Dirac fields 
and $\Delta^{(M)}=-\nabla^2\delta^\mu_\nu+R^\mu{}_\nu$ 
on Maxwell fields in the gauge $\alpha=1$.
Then, the ERGE is simply
\begin{equation*}
\partial_t\Gamma_k=\sum_{A=S,D,M}
\frac{n_A}{2}{\rm STr}_{(A)}\left(\frac{\partial_t P_k}{P_k}\right)
-{n_M}{\rm Tr}_{(S)}\left(\frac{\partial_t P_k}{P_k}\right)\ ,
\eqno(1.4.3)
\end{equation*}
where ${\rm STr}=\pm{\rm Tr}$ depending on the statistics,
and the last term comes from the ghosts.
Using integral transforms and the heat kernel expansion, 
the trace of a function $f$ of $\Delta$ can be expanded as
\begin{equation*}
{\rm Tr}f(\Delta)=
\sum_{n=0}^\infty Q_{2-n}(f) B_{2n}(\Delta) 
\eqno(1.4.4)
\end{equation*}
where the heat kernel coefficients $B_{2n}(\Delta)$ are linear combinations 
of the $\calo^{(n)}_i$, $Q_n(f)=(-1)^n f^{(n)}(0)$ for $n\leq0$
and $Q_n(f)$ are given by Mellin transforms of $f$ for $n>0$
\footnote{This technique is used also in some noncommutative geometry models,
see Chamseddine \& Connes (1996).}.
In this way one can write out explicitly the r.h.s. of (1.4.3) in terms of 
the $\calo^{(n)}_i$ and read off the beta functions.

When $N\to\infty$, this is the dominant contribution to the
gravitational beta functions, and graviton loops can be neglected 
(Tomboulis (1977), Smolin (1982), Percacci (2005)).
The functions $a^{(n)}_i$ defined in (1.2.5) 
become numbers;
with the so--called optimized cutoff function 
$R_k(z)=(k^2-z)\theta(k^2-z)$, discussed in Litim (2001, 2004), they are
\begin{align*}
a^{(0)}=&\frac{n_S-4n_D+2n_M}{32\pi^2}\ ,
\qquad a^{(1)}=\frac{n_S+2n_D-4n_M}{96\pi^2}\ ,\\
a^{(2)}_1=&\frac{6 n_S+36 n_D+72 n_M}{11520\pi^2}\ ,
\qquad a^{(2)}_2=\frac{10 n_S}{11520\pi^2}\ ,
\end{align*}
while $a^{(n)}_i=0$ for $n\geq 3$.
The beta functions (1.2.5) are then
\begin{equation*}
\partial_t \tilde g^{(n)}_i=(2n-4)\tilde g^{(n)}_i+a^{(n)}_i\ .
\eqno(1.4.5)
\end{equation*}
For $n\not= 2$ this leads to a FP
$$
\tilde g^{(n)}_{i*}=\frac{a^{(n)}_i}{4-2n}\ ,
\eqno(1.4.6)
$$
in particular we get
\begin{equation*}
\tilde\Lambda_*=-\frac{3}{4}\frac{n_S-4n_D+2 n_M}{n_S+2n_D-4n_M}\ ,\ \
\tilde G_*=\frac{12\pi}{-n_S-2n_D+4n_M}\ .
\eqno(1.4.7)
\end{equation*}
For $n=2$, one gets instead 
$g^{(2)}_i(k)=g^{(2)}_i(k_0)+a^{(2)}_i\ln(k/k_0)$,
implying asymptotic freedom for the couplings $\lambda$ and $\xi$ of (1.3.2). 
Remarkably, with this cutoff all the higher terms are zero at the FP.
The critical exponents are equal to the canonical dimensions of the $g^{(n)}$'s,
so $\Lambda$ and $G$ are UV--relevant (attractive),
$\lambda$ and $\xi$ are marginal and all the higher terms are UV--irrelevant.
Note that in perturbation theory $G$ would be UV--irrelevant (nonrenormalizable).
At the nontrivial FP the quantum corrections conspire with the classical
dimensions of $\Lambda$ and $G$ to reconstruct the dimensions of $g^{(0)}$ and $g^{(1)}$.
This does not happen at the Gau\ss ian FP, where the transformation between
$\tilde G$ and $\tilde g^{(1)}$ is singular.

\begin{figure}  
{\includegraphics{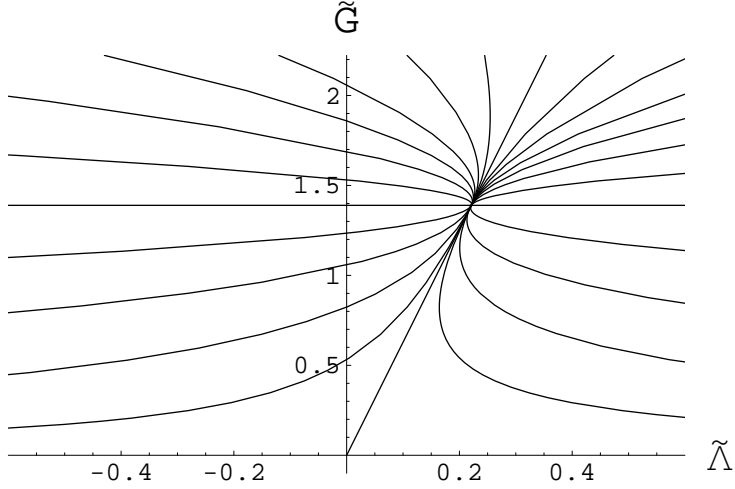}}
 \caption{The flow in the upper $\tilde\Lambda$--$\tilde G$ plane for pure gravity with
higher derivative terms at one loop, eq.(1.4.8). All other couplings are set to zero. 
The nontrivial FP at (0.221,1.389) is UV--attractive with eigenvalues $(-4,-2)$, 
the one in the origin is UV--attractive along the $\tilde\Lambda$ axis with eigenvalue $-2$
and repulsive in the direction of the vector $(1/2\pi,1)$ with eigenvalue $2$.}
\end{figure}

Using the same techniques, the one loop beta functions for gravity 
with the action (1.3.2) have been calculated by Codello \& Percacci (2006).
The beta functions for $\lambda$ and $\xi$ agree with those
derived in the earlier literature on higher derivative gravity
(Fradkin \& Tseytlin (1982), Avramidy \& Barvinsky (1985),
de Berredo--Peixoto \& Shapiro (2005)).
These couplings tend logarithmically to zero
with a fixed ratio $\omega=-3\lambda/\xi\to \omega_*=-0.023$.
The beta functions of $\tilde\Lambda$ and $\tilde G$
differ from the ones that were given in the earlier literature
essentially by the first two terms of the expansion (1.4.4).
In a conventional calculation of the effective action these terms 
would correspond to quartic and quadratic divergences, 
which are normally neglected in dimensional regularization,
but are crucial in generating a nontrivial FP.
Setting the dimensionless couplings to their FP--values, one obtains:
\begin{equation*}
\beta_{\tilde\Lambda} = 2\tilde\Lambda+\frac{2\tilde G}{\pi}-q_*\tilde G \tilde\Lambda\ ,\qquad \qquad 
\beta_{\tilde G}=2\tilde G-q_* \tilde G^2\ .
\eqno(1.4.8)
\end{equation*}
where $q_*\approx 1.440$.
This flow is qualitatively identical to the flow in the $N\to\infty$ limit,
and is shown in fig.1.

In order to appreciate the full nonperturbative content of the ERGE,
let us consider pure gravity in the Einstein--Hilbert truncation,
\ie\ neglecting terms with $n\geq 2$.
In a suitable gauge the operator 
$\frac{\delta^2\Gamma_k}{\delta g_{\mu\nu}\delta g_{\rho\sigma}}$ 
is a function of $-\nabla^2$ only.
Then, rather than taking as $\Delta$ the whole linearized wave operator,
as we did before, we use (1.4.4) with $\Delta=-\nabla^2$.
In this way we retain explicitly the dependence on $\Lambda$ and $R$.
Using the optimized cutoff, with gauge parameter $1/\alpha=Z$, the ERGE gives
\begin{align*}
\beta_{\tilde \Lambda}=&
\frac{-2(1-2\tilde\Lambda)^2\tilde\Lambda
+\frac{36-41\tilde\Lambda+42\tilde\Lambda^2-600\tilde\Lambda^3}{72\pi}\tilde G
+\frac{467-572\tilde\Lambda}{288\pi^2}\tilde G^2}
{(1-2\tilde\Lambda)^2-\frac{29-9\tilde\Lambda}{72\pi}\tilde G}\tag{1.4.9}\\
\beta_{\tilde G}=&  
\frac{2(1-2\tilde\Lambda)^2\tilde G
-\frac{373-654\tilde\Lambda+600\tilde\Lambda^2}{72\pi}\tilde G^2}
{(1-2\tilde\Lambda)^2-\frac{29-9\tilde\Lambda}{72\pi}\tilde G}\tag{1.4.10}
\end{align*}
This flow is shown in Figure 2.

\begin{figure}  
{\includegraphics{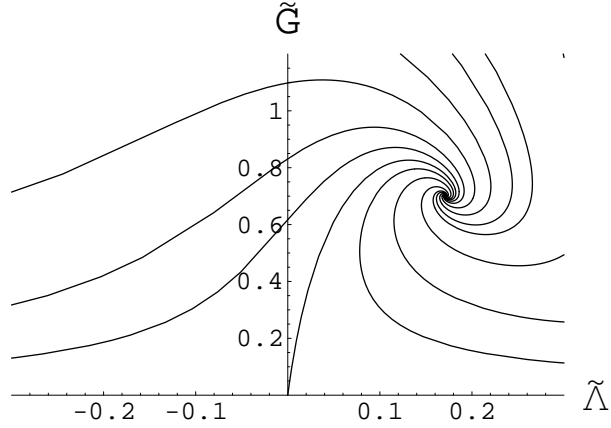}}
 \caption{The flow in the Einstein--Hilbert truncation, see Eq.(1.4.9-10). 
The nontrivial FP at $\tilde\Lambda=0.171$, $\tilde G=0.701$ is UV--attractive with eigenvalues $-1.69\pm 2.49i$. The Gau\ss ian FP is attractive along the $\tilde\Lambda$--axis
with eigenvalue $-2$ and repulsive in the direction $(0.04,1.00)$ with eigenvalue $2$.}
\end{figure}

Lauscher \& Reuter (2002a), Reuter \& Saueressig (2002) have studied 
the gauge-- and cutoff--dependence of the FP in the Einstein--Hilbert truncation. 
The dimensionless quantity $\Lambda'=\Lambda G$ (the cosmological constant
in Planck units) and the critical
exponents have a reassuringly weak dependence on these parameters.
This has been taken as a sign that the FP is not an artifact of the truncation.
Lauscher \& Reuter (2002b) have also studied the ERGE 
including a term $R^2$ in the truncation. 
They find that in the subspace of $\tilde\theories$ spanned by $\tilde\Lambda,\tilde G,1/\xi$, 
the non--Gau\ss ian FP is very close to the one of the Einstein--Hilbert truncation, 
and is UV--attractive in all three directions.
More recently, the FP has been shown to exist if the Lagrangian density is
a polynomial in $R$ of order up to six (Codello, Percacci and Rahmede (2008)).
In this truncation the UV critical surface is three dimensional.

There have been also other generalizations. 
Niedermaier (2003) considered the RG flow for dimensionally reduced $d=4$
gravity, under the hypothesis of the existence of two Killing vectors.
This subsector of the theory is parametrized by infinitely many couplings, 
and has been proved to be asymptotically safe.

Matter couplings have been considered by
Percacci \& Perini (2003a,b). Consider the general action
$$
\Gamma_k(g_{\mu\nu},\phi)=\int d^4x\,\sqrt{g}
\left(-\frac{1}{2}g^{\mu\nu}\partial_\mu\phi\partial_\nu\phi
-V(\phi^2)+F(\phi^2)R\right)\ ,
\eqno(1.4.11)
$$
where $V$ and $F$ are arbitrary functions of $\phi^2$, analytic at $\phi^2=0$.
This action has a so-called Gau\ss ian--Matter FP, meaning that only the coefficients
of the $\phi$-independent terms in (1.4.11) (namely $g^{(0)}$ and $g^{(1)}$) are nonzero.
The critical surface has dimension four and there are no marginal operators.
In the presence of other, minimally coupled matter fields, 
the dimension of the critical surface can be larger, and
it is easy to find theories where a polynomial potential
in $\phi$ is renormalizable and asymptotically free.
Thus, gravity seems to provide a solution to the so--called triviality problem
of scalar field theory.

It is tempting to speculate with Fradkin \& Tseytlin (1982)
that in the presence of gravity
all matter interactions are asymptotically free.
One loop calculations reported in Buchbinder \etal\ (1992), Robinson \& Wilczek (2005)
indicate that this may be the case also for gauge and Yukawa interactions.
Then, in studying the FP, it would be consistent to neglect matter 
interactions, as we did in the $1/N$ expansion.
If this is the case, it may become possible to show asymptotic safety
for realistic unified theories including gravity and the SM.

For the time being, the gravitational FP has been found
with a number of different approximations:
the $2+\epsilon$ expansion, the $1/N$ expansion,
polynomial truncations with a variety of cutoffs and gauges,
the two Killing vector reduction
and the most general four--derivative gravity theory at one loop.
The fact that all these methods yield broadly consistent results
should leave little doubt about the existence of a nontrivial 
FP with the desired properties.

\medskip
\section{Other approaches and applications}
\smallskip
In this final section we briefly comment on the relation of asymptotic safety
to other approaches and results in quantum gravity.

Gravity with the Einstein--Hilbert action has been shown by Goroff \& Sagnotti (1986)
and van de Ven (1992) to be perturbatively nonrenormalizable at two loops.
Stelle (1977) proved that the theory with action (1.3.2) and $\Lambda=0$ 
is perturbatively renormalizable:
all divergences can be absorbed into redefinitions of the couplings.
In general, asymptotic safety does not imply that in the UV limit 
only a finite number of terms in (1.3.1) survive:
there could be infinitely many terms, but there would be relations between their 
coefficients in such a way that only a finite number of parameters would be left free.
At one loop or in the large--$N$ limit, the ERGE predicts
that the UV critical surface can be parametrized by
the four couplings $\tilde\Lambda$, $\tilde G$, $\lambda$ and $\xi$,
the first two being nonzero at the FP and UV--relevant, the latter two being
asymptotically free and marginal.
Thus, at least in some approximations, asymptotic safety implies 
that near the FP quantum corrections to the action (1.3.2) will not generate 
new terms when one takes the UV limit.
This is very similar to the result of Stelle.
The main difference lies therein, that the perturbative proof
holds at the Gau\ss ian FP while the statement of asymptotic safety 
holds near the non--Gau\ss ian one.
According to the ERGE, the Gau\ss ian FP is unstable, and moving
by an infinitesimal amount towards positive $\tilde G$
(even with $\tilde\Lambda=0$)
would cause the system to be dragged in the direction of the repulsive
eigenvector towards the non-Gau\ss ian FP (see fig.1).
It is unclear whether in a more accurate description it will still
be possible to describe the UV limit of the theory by an action containing 
finitely many terms.

We now come to other nonperturbative approaches to quantum gravity.
Monte Carlo simulations of quantum gravity 
have found evidence of a phase transition which can be related
to the existence of a gravitational FP.
Hamber \& Williams (2004) review various results and arguments,
mainly from quantum Regge calculus,
supporting the claim that the mass critical exponent $\nu$ is equal to 1/3. 
In a theory with a single coupling constant $\tilde G$
we have $-1/\nu=\beta'_{\tilde G}(\tilde G_*)$, so for a rough comparison we can
solve (1.4.10) with $\tilde\Lambda=0$, finding a FP at $\tilde G_*=1.21$
with $\beta'(\tilde G_*)\approx -2.37$.
The agreement is numerically not very good for a universal quantity,
but it might perhaps be improved by taking into account the flow 
of the cosmological constant.

In the so--called causal dynamical triangulation approach, 
recent numerical simulations have produced quantum worlds that exhibit
several features of macroscopic four--dimensional spacetimes 
(see Ambj\o rn, Jurkiewicz and Loll's contribution to this volume).
In particular they have also studied diffusion processes in such quantum spacetimes
and found that the spectral dimension characterizing them
is close to two for short diffusion times and to four for long diffusion times.
This agrees with the expectation from asymptotic safety and can be seen as further
independent evidence for a gravitational FP, as we shall mention below.
																		
The physical implications of a gravitational FP and, more generally, of the
running of gravitational couplings, are not yet well understood.
First and foremost, one would expect asymptotic safety to lead to new insight into the local, 
short--distance structure of a region of spacetime.
The boundedness of the cutoff in Planck units, derived in section 1.3, 
would be in accord with the widely held expectation of some kind of discrete spacetime 
structure at a fundamental level.
In particular, it may help understand the connection
to theories such as loop quantum gravity, which predict that areas
and volumes have quantized values.
However, the discussion in section 1.3 should make it clear
that the issue of a minimal length in quantum gravity
may have limited physical relevance, 
since the answer depends on the choice of units.

Another point that seems to emerge is that the spacetime geometry cannot be understood in
terms of a single metric: rather, there will be a different effective metric
at each momentum scale.
This had been suggested by Floreanini \& Percacci (1995a,b),
who calculated the scale dependence of the metric using an effective
potential for the conformal factor.
Such a potential will be present in the effective action $\Gamma_k$ before the 
background metric is identified with the classical metric (as mentioned in section 1.2).
A scale dependence of the metric has also been postulated by Magueijo \& Smolin (2004)
in view of possible phenomenological effects.
Lauscher \& Reuter (2005) have suggested the following picture of a fractal spacetime.
Dimensional analysis implies that in the FP regime
$\langle g_{\mu\nu}\rangle_k=k^{-2}(\tilde g_0)_{\mu\nu}$, where $\tilde g_0$,
defined as in (1.3.4),
is a fiducial dimensionless metric that solves the equations of motion of $\Gamma_{k_0}$.
For example, in the Einstein--Hilbert truncation, the effective metric $\langle\gmn\rangle_k$
is a solution of the equation $R_{\mu\nu}=\Lambda_k g_{\mu\nu}$, so
$$
\langle g_{\mu\nu}\rangle_k=
\frac{\Lambda_{k_0}}{\Lambda_{k}}\langle g_{\mu\nu}\rangle_{k_0}
\approx \left(\frac{k_0}{k}\right)^2\langle g_{\mu\nu}\rangle_{k_0}=k^{-2}(\tilde g_0)_{\mu\nu}\ ,
\eqno(1.5.1)
$$
where $\approx$ means ``in the FP regime''.
The fractal spacetime is described by the collection of all these metrics.

A phenomenon characterized by an energy scale $k$ will ``see''
the effective metric $\langle g_{\mu\nu}\rangle_k$.
For a (generally off-shell) free particle with four--momentum $p_\mu$ 
it is natural to use $k\propto p$, where $p=\sqrt{(\tilde g_0)^{\mu\nu}p_\mu p_\nu}$.
Its inverse propagator is then $\langle  g^{\mu\nu}\rangle_{p}p_\mu p_\nu$.
At low energy $\langle g_{\mu\nu}\rangle_k$ does not
depend on $k$ and the propagator has the usual $p^{-2}$ behaviour;
in the FP regime, (1.5.1) implies instead that it is proportional to $p^{-4}$.
Its Fourier transform has a short--distance
logarithmic behaviour which is characteristic of two dimensions,
and agrees with the aforementioned numerical results on the spectral
dimension in causal dynamical triangulations.
This agreement is encouraging, because it suggests that the two approaches are 
really describing the same physics.
When applied to gravitons in four dimensions (and only in four dimensions!) 
it also agrees with the general prediction, derived in the end of section 1.3,
that $\eta_g=2$ at a nontrivial gravitational FP.

The presence of higher derivative terms in the FP action
raises the old issue of unitarity:
as is well--known, the action (1.3.2) describes, besides a massless graviton,
also particles with Planck mass and negative residue (ghosts).
From a Wilsonian perspective, this is clearly not very significant:
to establish the presence of a propagator pole at the mass $m_P$ one should
consider the effective action $\Gamma_k$ for $k\approx m_P$,
which may be quite different from the FP action.
Something of this sort is known to happen in the theory of strong interactions: 
at high energy they are described by a renormalizable and asymptotically free theory
(QCD), whose action near the UV (Gau\ss ian) FP describes quarks and gluons.
Still, none of these particles appears in the physical spectrum.

As in QCD, matching the UV description to low energy phenomena
may turn out to be a highly nontrivial issue.
A change of degrees of freedom could be involved.
From this point of view one should not assume {\it a priori} that the metric appearing 
in the FP action is ``the same'' metric that appears in the low energy description of GR.
Aside from a field rescaling, as discussed in section 1.2,
a more complicated functional field redefinition may be necessary,
perhaps involving the matter fields, as exemplified in Tomboulis (1996).
Unless at some scale the theory was purely topological, it will always involve a metric
and from general covariance arguments it will almost unavoidably contain an 
Einstein--Hilbert term.
This explains why the Einstein--Hilbert action, which describes GR at macroscopic distances,
may play an important role also in the UV limit, as the results of section 1.4 indicate.
With this in mind, one can explore the consequences
of a RG running of gravitational couplings also in other regimes.

Motivated in part by possible applications to the hierarchy problem,
Percacci (2007) considered a theory with an action of the form (1.4.11),
in the intermediate regime between the scalar mass and the Planck mass.
Working in cutoff units (1.3.4), it was shown that the warped geometry
of the Randall--Sundrum model can be seen as a geometrical
manifestation of the quadratic running of the mass.

For applications to black hole physics,
Bonanno \& Reuter (2000) have included quantum gravity effects 
by substituting $G$ with $G(k)$ in the Schwarzschild metric, 
where $k=1/r$ and $r$ is the proper distance from the origin.
This is a gravitational analog of the \"Uhling approximation of QED.
There is a softening of the singularity at $r=0$, and it is predicted
that the Hawking temperature goes to zero for Planck mass black holes, 
so that the evaporation stops at that point.

In a cosmological context, it would be natural to identify the scale $k$
with a function of the cosmic time.
Then, in order to take into account the RG evolution of the couplings,
Newton's constant and the cosmological constant can be replaced 
in Friedman's equations by the effective Newton's constant 
and the effective cosmological constant calculated from the RG flow.
With the identification $k=1/t$, where $t$ is cosmic time,
Bonanno \& Reuter (2002) have applied this idea to the Planck era,
finding significant modifications to the cosmological evolution; 
a more complete picture extending over all of cosmic history
has been given in Reuter \& Saueressig (2005).
It has also been suggested that an RG running of gravitational
couplings may be responsible for several astrophysical or cosmological effects.
There is clearly scope for various interesting speculations, 
which may even become testable against new cosmological data.

Returning to the UV limit, it can be said that 
asymptotic safety has so far received relatively little attention,
when compared to other approaches to quantum gravity.
Establishing this property is obviously only the first step:
deriving testable consequences is equally important and may prove an even 
greater challenge.
Ultimately, one may hope that asymptotic safety will play a similar role
in the development of a QFT of gravity as asymptotic freedom played 
in the development of QCD.

\section{Acknowledgements}
I wish to thank R. Floreanini, D. Dou, D. Perini, A. Codello and C. Rahmede
for past and present collaborations, and M. Reuter for many discussions.
\vfill
\break

\break

\centerline{\bf Questions and answers}
\bigskip

{\bf Q}: Could an asymptotically safe theory be regarded as an
approximation to another more fundamental theory, or does it have
to be regarded as a self-contained fundamental theory?

{\bf A}: The asymptotic safety programme is very closely related to the
formalism of effective field theories and both possibilities
can be envisaged.
If a fixed point with the desired properties did exist, then mathematically
it would be possible to take the limit $k\to\infty$ and one could call this
a fundamental theory.
It would do for gravity what the Weinberg--Salam model originally did
for electroweak interactions.
However, experience shows that today's fundamental theory 
may become tomorrow' effective theory.
The renormalizability of the Weinberg--Salam model was important in establishing 
it as a viable theory but nowadays this model is widely regarded 
as an effective theory whose nonrenormalizable couplings are suppressed 
by powers of momentum over some cutoff. In a distant future, the same could 
happen to an asymptotically safe theory of gravity.

To understand this point better, notice that
in order to hit the fixed point as $k\to\infty$, 
one would have to place the initial point of the flow
in the critical surface with ``infinite precision''.
In the case of the standard model, where the use of
perturbative methods is justified, this corresponds to setting
all couplings with negative mass dimension {\it exactly} equal to zero.
Even assuming that the property of asymptotic safety could be firmly
established theoretically, because measurements are always imprecise,
it is hard to see how one could ever establish 
experimentally that the world is described by such a theory.
One could say at most that experiments are compatible with the theory
being fundamental.

On the other hand suppose that the theory requires drastic modification
at an energy scale of, say, a billion Planck masses, perhaps because
of the existence of some presently unknown interaction.
Then at the Planck scale one would expect the dimensionless couplings 
of the theory ($\tilde g_i$) to lie off the critical surface by an amount 
of the order of some power of one in a billion.
Suppose we follow the flow in the direction of decreasing energies
starting from a scale which is much larger than one, and much less
than a billion Planck masses.
Since the fixed point is IR--attractive in all directions except
the ones in the critical surface, starting from a generic point in the space 
of coupling constants, the theory will be drawn quickly towards the critical surface.
Going towards the infrared, the flow at sub--Planckian scales 
will then look as if it had originated from the fixed point, 
up to small deviations from the critical surface
which may be hard or impossible to measure.

Thus, the formalism can accommodate both effective and
fundamental theories of gravity. The most important point is that
asymptotic safety would allow us to push QFT beyond the Planck scale,
up to the next frontier, wherever that may be.

\medskip
{\bf Q}: What is your take on the issue of continuum versus discrete picture of
spacetime, coming from a renormalization group perspective? If gravity is
asymptotically safe, would it imply that a continuum description of
spacetime is applicable at all scales, or one can envisage a role of
discrete spacetime structures even in this case? How would a breakdown of
the continuum description show up in the ERG approach?

{\bf A}: First of all it should be said that the renormalization group can be
realized both in continuum and discrete formulations and is likely to play a role
in quantum gravity in either case. It should describe the transition from
physics at the ``lattice'' or UV cutoff scale down to low energies.

Then, one has to bear in mind that when one formulates a quantum field theory 
in the continuum but with a cutoff $\Lambda$, it is impossible to resolve 
points closer that $1/\Lambda$, so the continuum should be regarded as a 
convenient kinematical framework that is devoid of physical reality.
If the asymptotic safety program could be carried through literally as described,
it would provide a consistent description of physics down to arbitrarily
short length scales, and in this sense the continuum would become, at least
theoretically, a reality.

Of course, it would be impossible to establish experimentally the continuity of spacetime 
in the mathematical sense, so this is not a well--posed physical question.
What is in principle a meaningful physical question, and may become answerable
sometimes in the future, is whether spacetime is continuous down to, say, 
one tenth of the Planck length.
But even then, the answer may require further qualification.
Recall that in order to define a distance one has to specify a unit of lengths.
Units can ultimately be traced to some combination of
the couplings appearing in the action. For example, in Planck units one takes
the square root of Newton's constant as a unit of length.
Because the couplings run, when the cutoff is sent to infinity the distance 
between two given points could go to zero, to a finite limit or to infinity 
depending on the asymptotic behaviour of the unit.
In principle it seems possible that spacetime looks discrete in certain
units and continuous in others.
Then, even if asymptotic safety was correct, it need not be in conflict
with models where spacetime is discrete.

\medskip
{\bf Q}: What differences, in formalism and results, can one expect in the ERG
approach, if one adopts a 1st order (e.g. Palatini) or BF-like (e.g.
Plebanski) description of gravity?

{\bf A}: Writing the connection as the sum of the Levi--Civita connection and
a three-index tensor $\Phi$, one can always decompose an action for independent
connection and metric into the same action written for the Levi-Civita connection,
plus terms involving $\Phi$. The effects due to $\Phi$ will be similar to those
of a matter field.
In the case when the action is linear in curvature, and possibly quadratic in torsion
and nonmetricity, up to a surface term the action for $\Phi$ is just a mass term, 
implying that $\Phi$ vanishes on shell.
In this case one expects the flow to be essentially equivalent to that obtained in the
Einstein-Hilbert truncation plus some matter fields, although this has not been explicitly checked yet.
The presence of a mass for $\Phi$ of the order of the Planck mass
suggests that a decoupling theorem is at work and that $\Phi$ (or equivalently the connection)
will become propagating degrees of freedom at the Planck scale.
This is indeed the case when the action involves terms quadratic in curvature
(which can be neglected at low energies).
Then the field $\Phi$ propagates, and has quartic self-interactions.
There will be new couplings, that may influence the running of Newton's constant,
for example. But again, this should be equivalent to fourth-order
gravity plus matter. 

\medskip
{\bf Q}: You mention that the results of the ERG seem to point out that spacetime
structure cannot be described in terms of a single metric for any momentum
scale. How would one notice, in the RG approach, that it cannot be described
by a metric field at all, but that a description in terms of connections or
even a non-local one would be more appropriate, say, at the Planck scale?

{\bf A}: I do in fact expect that an independent connection will manifest itself at the
Planck scale, as I have indicated in my answer to another question, though
I don't think that this will be forced upon us by the ERG.

The scale-dependence of the metric could manifest itself as violations 
of the equivalence principle, or perhaps as Lorentz-invariance
violations or deformations of the Lorentz group.
There is much work to be done to understand this type of phenomenology.
Even more radically, it is possible that gravity is just the ``low energy''
manifestation of some completely different physics,
as suggested in the article by Dreyer. This would probably imply a failure 
of the asymptotic safety programme, for example a failure to find a
fixed point when certain couplings are considered.

\medskip

{\bf Q}: Can you please comment on possibility of extending the ERG approach to
the Lorentzian signature or to the case of dynamical space topology?

{\bf A}: So far the ERG has been applied to gravity in conjunction with the background
field method. Calculations are often performed in a convenient background,
such as (Euclidean) de Sitter space, but the beta functions obtained in this way
are then completely general and independent of the background metric and
spacetime topology. The choice of a background is merely a calculational trick.
It is assumed that the beta functions are also independent of the signature of the
background metric, although this point may require further justification.
One should also stress in this connection that the use of the background field
method and of the background field gauge does not make this a 
``background-dependent'' approach. On the contrary, when properly implemented
it guarantees that the results are background-independent.


\begin{thereferences}{widest citation in source list}

\bibitem{keyX}
T. Aida and Y. Kitazawa (1997).
Two--loop prediction for scaling exponents in (2+$\epsilon$)--dimensional quantum gravity.
{\it  Nucl. Phys.}  {\bf B 491}, 427. 

\bibitem{keyX} 
I.G. Avramidy and A.O. Barvinsky (1985).
Asymptotic freedom in higher--derivative quantum gravity.
{\it  Phys. Lett.}  {\bf 159B}, 269. 

\bibitem{keyX} 
C. Bagnuls and C. Bervillier (2001).
Exact renormalization group equations: An introductory review.
{\it  Phys. Rept.}  {\bf 348}, 91. 

\bibitem{keyX} 
J. Berges, N. Tetradis, and C. Wetterich (2002).
Nonperturbative renormalization flow in quantum field theory and statistical physics.
{\it Phys. Rept.} {\bf 363}, 223.

\bibitem{keyX} 
N.E.J Bjerrum-Bohr, J.F. Donoghue and B.R. Holstein (2003).
Quantum gravitational corrections to the nonrelativistic scattering potential of two masses.
{\it Phys. Rev.} {\bf D 67}, 084033
[Erratum-ibid.\  {\bf D 71} (2005) 069903]

\bibitem{keyX}
A. Bonanno and M. Reuter (2000).
Renormalization group improved black hole spacetimes.
{\it  Phys. Rev.} {\bf D 62}, 043008.

\bibitem{keyX}
A. Bonanno and M. Reuter (2002).
Cosmology of the Planck era from a renormalization group for quantum gravity.
{\it  Phys. Rev.} {\bf D 65}, 043508.

\bibitem{keyX}
I.L. Buchbinder, S.D. Odintsov and I. Shapiro (1992).
Effective action in quantum gravity.
IOPP Publishing, Bristol.

\bibitem{keyX}
A. Chamseddine and A. Connes (1996). 
The Spectral action principle.
Commun.Math.Phys.{\bf 186}, 731-750.

\bibitem{keyX}
S.M. Christensen and M.J. Duff (1978).
Quantum Gravity In Two + Epsilon Dimensions.
{\it  Phys. Lett.} {\bf B 79}, 213.

\bibitem{keyX}
A. Codello and R. Percacci (2006). 
Fixed points of higher derivative gravity.
Phys.Rev.Lett.97:221301

\bibitem{keyX}
A. Codello, R. Percacci and C. Rahmede (2007).  
Ultraviolet properties of f(R)-gravity,
Int. J. Mod. Phys. A23:143-150.

\bibitem{keyX} 
G. de Berredo--Peixoto and I.L. Shapiro (2005). 
Higher derivative quantum gravity with Gauss --Bonnet term.
{\it  Phys. Rev.} {\bf D 71}, 064005.

\bibitem{keyX} 
D. Dou and R. Percacci (1998). 
The running gravitational couplings. 
{\it Class. Quant. Grav.} {\bf 15}, 3449 

\bibitem{keyX} 
R. Floreanini and R. Percacci (1995a). 
Average effective potential for the conformal factor.
{\it Nucl. Phys.} {\bf B 436}, 141-160.

\bibitem{keyX} 
R. Floreanini and R. Percacci (1995b).
Renormalization--group flow of the dilaton potential.
{\it Phys. Rev.} {\bf D 52}, 896-911.

\bibitem{keyX} 
E.S. Fradkin and A.A. Tseytlin (1982).
Renormalizable asymptotically free quantum theory of gravity.
{Nucl. Phys.} {\bf B 201}, 469.

\bibitem{keyX}
R. Gastmans, R. Kallosh and C. Truffin (1978).
Quantum Gravity Near Two-Dimensions,
{\it  Nucl.\ Phys.} {\bf B 133}, 417.

\bibitem{keyX}
M.H. Goroff and A. Sagnotti (1986).
The Ultraviolet Behavior of Einstein Gravity.
Nucl.Phys.{\bf B266}, 709.

\bibitem{keyX}
H. Hamber and R. Williams (2004).
Non--perturbative gravity and the spin of the lattice graviton.
{\it  Phys. Rev.} {\bf D 70}, 124007.

\bibitem{keyX}
H. Kawai and M. Ninomiya (1990). 
Renormalization group and quantum gravity.
{\it Nuclear Physics} {\bf B 336}, 115-145.

\bibitem{keyX}
D.I. Kazakov (2003). 
Ultraviolet fixed points in gauge and SUSY field theories in extra dimensions.
{\it JHEP} {\bf 03}, 020.

\bibitem{keyX} 
O. Lauscher and M. Reuter (2002a). 
Ultraviolet fixed point and generalized flow equation of quantum gravity.
Phys. Rev. {\bf D65} 025013. 

\bibitem{keyX} 
O. Lauscher and M. Reuter (2002b).
Flow equation of quantum Einstein gravity in a higher derivative truncation.
{\it Phys. Rev.} {\bf D 66}, 025026. 

\bibitem{keyX} 
O. Lauscher and M. Reuter (2005).
Fractal spacetime structure in asymptotically safe gravity.
{\it JHEP} {\bf 0510}, 050. 

\bibitem{keyX}
D.F. Litim (2001). 
Optimised renormalisation group flows.
{\it  Phys. Rev.}  {\bf D 64}, 105007.

\bibitem{keyX}
D. Litim (2004).
Fixed points of quantum gravity.
{\it Phys. Rev. Lett.} {\bf 92}, 201301.

\bibitem{keyX} 
J. Magueijo and L. Smolin (2004). 
Gravity's rainbow.
{\it  Class. and Quantum Grav.}  {\bf 21}, 1725.

\bibitem{keyX} 
M. Niedermaier (2003).
Dimensionally reduced gravity theories are asymptotically safe.
{\it Nucl. Phys.} {\bf B 673}, 131-169.

\bibitem{keyX} 
M. Niedermaier and M. Reuter (2006).
The Asymptotic Safety Scenario in Quantum  Gravity.
Living Rev. Relativity 9,  (2006),  5. 

\bibitem{keyX}
J.M. Pawlowski (2005). 
Aspects of the functional renormalization group.
[arXiv:hep-th/0512261].

\bibitem{keyX} 
R. Percacci and D. Perini (2003a).
Constraints on matter from asymptotic safety.
{\it Phys. Rev.} {\bf D67}, 081503 (R).

\bibitem{keyX} 
R. Percacci and D. Perini (2003b).
Asymptotic safety of gravity coupled to matter.
{\it Phys. Rev.} {\bf D68}, 044018 .

\bibitem{keyX} 
R. Percacci and D. Perini (2004).
On the ultraviolet behaviour of Newton's constant.
{\it  Class. and Quantum Grav.} {\bf 21}, 5035.

\bibitem{keyX} 
R. Percacci (2006). 
Further evidence for a gravitational fixed point.
{\it Phys. Rev.} {\bf D73}, 041501(R).

\bibitem{keyX} 
R. Percacci (2007). 
The renormalization group, systems of units and the hierarchy problem.
J.Phys. {\bf A40}, 4895-4914.

\bibitem{keyX} 
M. Reuter (1998). 
Nonperturbative evolution equation for quantum gravity.
{\it Phys. Rev.} {\bf D57}, 971.

\bibitem{keyX} 
M. Reuter and F. Saueressig (2002).
Renormalization group flow of quantum gravity in the Einstein--Hilbert truncation.
{\it Phys. Rev.} {\bf D65}, 065016.

\bibitem{keyX} 
M. Reuter and F. Saueressig (2005). 
From big bang to asymptotic de Sitter: Complete cosmologies in a quantum gravity framework.
{\it JCAP} {\bf 09}, 012.

\bibitem{keyX} 
S.P. Robinson and F. Wilczek (2005).
Gravitational corrections to running gauge couplings.
Phys.Rev.Lett.{\bf 96}, 231601

\bibitem{keyX}
L. Smolin (1982). 
A fixed point for quantum gravity.
{\it Nucl. Phys.} {\bf B 208}, 439-466.

\bibitem{keyX}
W. Souma (1999). 
Nontrivial ultraviolet fixed point in quantum gravity.
{\it Prog. Theor. Phys.} {\bf 102}, 181.

\bibitem{keyX}
K.S. Stelle (1977). 
Renormalization of higher--derivative gravity.
{\it Phys. Rev.} {\bf D 16}, 953-969.

\bibitem{keyX}
E. Tomboulis (1977). 
1/N expansion and renormalizability in quantum gravity
{\it Phys. Lett.} {\bf 70 B}, 361.

\bibitem{keyX}
E. Tomboulis (1996).
Exact relation between Einstein and quadratic quantum gravity.
{\it Phys. Lett.} {\bf B 389}, 225.

\bibitem{keyX}
A.E.M. van de Ven (1992).
Two loop quantum gravity.
{\it Nucl. Phys.} {\bf B 378}, 309-366.

\bibitem{keyX} 
S. Weinberg (1979). 
Ultraviolet divergences in quantum theories of gravitation. 
In {\it General Relativity: An Einstein centenary survey}, 
ed. S.~W. Hawking and W. Israel, chapter 16, pp.790--831; 
Cambridge University Press.

\bibitem{keyX}
C. Wetterich (1993).
Exact evolution equation for the effective potential.
{\it Phys. Lett.} {\bf B 301}, 90.

\end{thereferences}
\end{document}